\documentclass[a4paper,fleqn]{cas-dc}

\usepackage[numbers]{natbib}
\usepackage{xcolor}

\def\tsc#1{\csdef{#1}{\textsc{\lowercase{#1}}\xspace}}
\tsc{WGM}
\tsc{QE}
\tsc{EP}
\tsc{PMS}
\tsc{BEC}
\tsc{DE}
\usepackage{caption}
\begin{document}
	\let\WriteBookmarks\relax
	\def\floatpagepagefraction{1}
	\def\textpagefraction{.001}
	\shorttitle{}
	\shortauthors{S.Chen H.Liang J.Wang et al.}
	
	\title [mode = title]{Spin Faraday pattern formation in a circular spin-orbit coupled Bose-Einstein condensate with stripe phase}

	\author[1]{Shixiang Chen}[
                        style=chinese]

\credit{Conceptualization of this study, Methodology, Software}

\affiliation[1]{organization={School of Physics, East China Normal University},
                addressline={Dongchuan Road}, 
                city={Shanghai},
                postcode={200241}, 
                country={China}}

\author[2]{Hongguang Liang}[style=chinese]

\author[1]{Juan Wang}[style=chinese,
   ]

\credit{Data curation, Writing - Original draft preparation}

\affiliation[2]{organization={College of Electronic Information and Optical Engineering, Nankai University},
                addressline={Tongyan Road}, 
                postcode={300000}, 
                city={Tianjin},
                country={China}}

\affiliation[3]{organization={Chongqing Key Laboratory of Precision Optics, Chongqing Institute of East China Normal University},
                postcode={401120}, 
                city={Chongqing},
                country={China}}

\author[1,3]{Yan Li}[style=chinese,
                        orcid=0000-0003-1504-413X]
                        	\cormark[1]
	
	\ead{yli@phy.ecnu.edu.cn}

\cortext[cor1]{Corresponding author}

	\begin{abstract}
We investigate the spin Faraday pattern formation in a periodically driven, pancake-shaped spin-orbit-coupled (SOC) Bose-Einstein condensate (BEC) prepared with stripe phase. By modulating atomic interactions using in-phase and out-of-phase schemes, we observe collective excitation modes with distinct rotational symmetries (L-fold). Crucially, at the critical modulation frequency, out-of-phase modulation destabilizes the L = 6 pattern, whereas in-phase modulation not only preserves high symmetry but also excites higher-order modes. Unlike conventional binary BECs, Faraday patterns emerge here without initial noise due to SOC-induced symmetry breaking, with all patterns exhibiting supersolid characteristics. Furthermore, we demonstrate control over pattern symmetry, radial nodes, and pattern radius by tuning the modulation frequency, providing a new approach for manipulating quantum fluid dynamics. This work establishes a platform for exploring supersolidity and nonlinear excitations in SOC systems with stripe phase.

	\end{abstract}


	\maketitle
	
	\section{Introduction}
	 Pattern formation represents a fundamental phenomenon that reveals intrinsic properties of physical systems. In chemistry, it unveils molecular dynamics for reaction rate study \cite{bib10}; in cosmology, preheating-phase patterns reveal relativistic nonlinear effects \cite{bib7}; in materials science, domain patterns exhibit crystal stacking and interlayer interactions \cite{bib4}. A seminal contribution emerged in 1831 when Michael Faraday discovered standing wave patterns, now termed Faraday waves, through experiments on parametrically driven liquid surfaces \cite{bib11}. Since the discovery of stable standing waves formed in a vibrating fluid layer within a container, various patterns that form in fluids or quantum fluids have been revealed \cite{bib1,bib2,bib8,bib12}. 
    
  The advent of Bose-Einstein condensates (BEC) \cite{bib13} in 1995 shifted Faraday wave research toward quantum fluid systems. BEC offers exceptional tunability, enabling precise investigations of nonlinear excitations in both bosonic systems \cite{bib14,bib15,bib16,bib17} and fermionic systems \cite{wuhaibin,wenwen}. This has spurred extensive theoretical and experimental studies, with Faraday wave phenomena successfully demonstrated in single-component systems via periodic modulation of scattering lengths or confinement frequencies \cite{bib19,bib9,bib20,bib20.1}, and extended to two-component condensates \cite{bib21,bib23,bib44}. 
  
  Recent advances in spin-orbit-coupled (SOC) BEC have further expanded this paradigm \cite{bib24,bib25}. SOC mechanisms, crucial for phenomena like the spin Hall effect \cite{bib26,bib27} and topological insulators \cite{bib28,bib29}, introduce rich interplay between spin and momentum degrees of freedom. The SOC Bose system, a promising platform for exploring novel Faraday wave dynamics in quantum regimes, in which the stripe phase manifests itself through counter-propagating atomic momenta for spin-up and spin-down components. This momentum opposition induces stripe separation, spontaneously breaking the translational symmetry of the system \cite{bib33,bib34,bib35}, thereby establishing a prototypical platform for supersolid research \cite{bib30,bib31,bib32,supersolid4}. Experimentally, Bragg scattering techniques have enabled the observation and analysis of supersolid density modulations induced by SOC BEC \cite{bib36.5}.
    
  Faraday waves have been theoretically studied by quenching the strength of Raman coupling in SOC BEC system \cite{bib36}. However, research on Faraday waves generated via modulating the atomic interactions in SOC BEC systems are limited. In our earlier work, we investigated the Faraday waves in a normal binary BEC \cite{meiling2025} and in an elongated SOC BEC \cite{bib52} by modulating atomic interactions. Now we focus on the pancake shaped SOC BEC system with stripe phase through periodically modulating the atomic scattering length to explore more intriguing physics.
    
    In this paper, we periodically modulate the interatomic interactions via two different schemes with stripe phase of SOC BEC to generate patterns. All the patterns exhibit L-fold rotational symmetry, which means that the density distribution of each pattern will return to its initial position after a rotation by an angle of $2\pi/L$. Under $out$-$of$-$phase$ modulation, the rotational symmetry patterns are excited when L $<$ 6, and the pattern become unstable when L = 6, which is consistent with the results from the other systems \cite{bib41.1,bib41.2}. However, with $in$-$phase$ modulation, the Faraday patterns show high symmetry and higher-order rotational symmetry when L $\ge$ 6, which can not be easily excited in other systems \cite{bib41.1,bib41.2}. For single-component BEC \cite{bib41.2} and normal binary BEC with $in$-$phase$ modulation \cite{meiling2025}, an initial noise is necessary to excite Faraday patterns. However, such noise is not necessary due to the symmetry breaking in our SOC system. We find that under $out$-$of$-$phase$ modulation, the patterns and the dipole mode appear simultaneously, which results in the symmetry breaking. In contrast, under $in$-$phase$ modulation, the Faraday mode and dipole mode appear successively, leading to the high symmetry of Faraday patterns. With stripe phase, all the patterns exhibit supersolid characteristics, providing a platform for exploring supersolidity and nonlinear excitations in SOC systems.

	This paper is organized into four sections. Section II introduces the theoretical model of our periodically driven pancake-shaped SOC system with stripe phase. Section III systematically examines Faraday pattern excitation dynamics and the characteristics of the patterns under $in$-$phase$ and $out$-$of$-$phase$ modulations. Section IV is the conclusion.

	\section{Theoretical model}
	
	We use the scheme of I. B. Speileman to study the SOC BEC \cite{bib25}. We confine atoms in a harmonic potential $V=\frac{1}{2} m \omega_{x}^{2} x^{2}+\frac{1}{2} m \omega_{y}^{2} y^{2}+\frac{1}{2} m \omega_{z}^{2} z^{2}$, where $m$ is the atomic mass, and $\omega_{x}, \omega_{y}, \omega_{z}$ are the trap frequencies along the x-, y-, and z-directions. We select two hyperfine states of $^{87}\text{Rb}$ atoms as the pseudospin-up state $\left | \uparrow  \right \rangle$ = $\left  | F=1,m_F=0  \right \rangle$ and the pseudospin-down state $\left | \downarrow  \right \rangle$ = $\left  | F=1,m_F=-1  \right \rangle$. We select a pancake-shaped BEC with $\omega_{x} = \omega_{y} = 50$ Hz and $\omega_{z} = 1000$ Hz to ensure a quasi-2D system.

    The single-particle Hamiltonian of the SOC BEC under the rotating wave approximation is as follows \cite{bib42}:
	\begin{equation}
		H_{\mathrm{sp}}=\frac{1}{2 m}\left[\left(p_{x}-k_{r} \sigma_{z}\right)^{2}+p_{y}^{2}\right]+\frac{\hbar \Omega}{2} \sigma_{x}+\frac{\hbar}{2} \delta \sigma_{z}+V.\label{eq1}
	\end{equation}
	
	Here, $p_x$, $p_y$ are the momentum in the x- and y-directions. $k_r$ is the projected wavenumber of Raman laser along the counter propagating direction. $\delta $ represents the energy level difference between the two spin states. $\Omega $ represents the Raman coupling strength, which reflects the transition between the two energy levels. $\sigma_{x,z}$ denotes the Pauli matrices in the relevant directions.

\begin{figure}
		\centering
		\includegraphics[width=0.9\columnwidth]{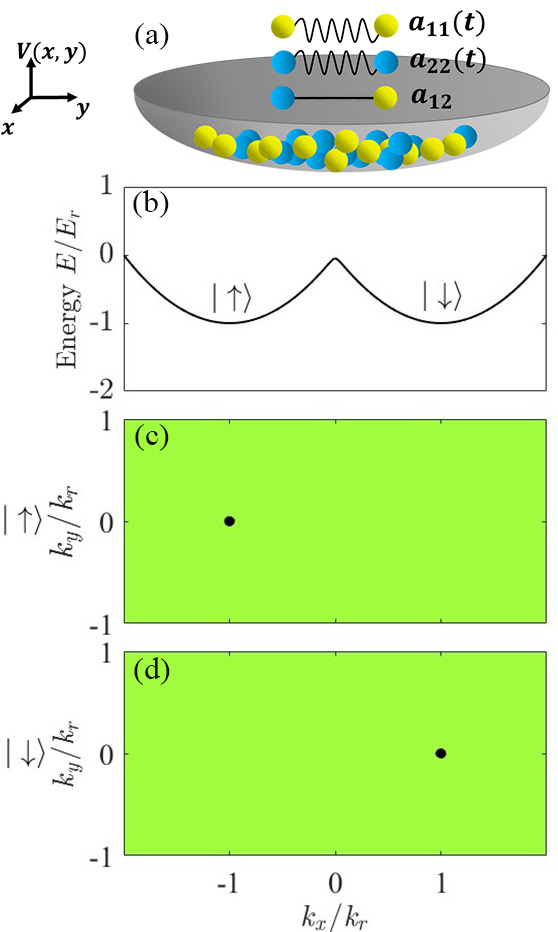}
		\caption{\rmfamily (a) Schematic of the harmonically trapped BEC with the atomic interaction periodically modulated. Here, 1 and 2 represent pseudo spin-up (yellow balls) and spin-down (blue balls). The scattering length $a_{11}$ and $a_{22}$ can be periodically modulated, while $a_{12}$ is a constant. (b) Dimensionless dispersion relation of stripe phase with $\Omega=0.1E_r,\delta=0$. The two minima correspond to the spin states $\left |\uparrow \right \rangle $ and $\left |\downarrow \right \rangle$. (c)(d) The momentum distributions of the two spin states in the x-direction are opposite.}
		\label{fig_theory}
	\end{figure}
	



We aim to study how atomic interactions influence the pattern formation of SOC system. The interaction of atoms is given by $g_{ij}={2\sqrt{2\pi} \hbar^2a_{ij}}/{ma_z}$, where $a_{ij}$ denotes the $s$-wave scattering length, and $a_z=\sqrt{\hbar/m\omega_z}$ is the harmonic oscillator length. Since the atomic interactions can be tuned by modulating the scattering lengths \cite{bib38, bib40, PhysRevA.51.4852, PhysRevLett.81.938, PhysRevLett.104.153201}, we apply periodical modulation to the scattering lengths by using two theoretical modulation protocols:
        	\begin{equation}
            \begin{cases}
a_{11}(t) = a_{11} + A\cos(\omega_m t) \\
a_{22}(t) = a_{22} \pm A\cos(\omega_m t)
\end{cases}
		.\label{eq7}
	\end{equation}
    
In \textcolor{blue}{Eq.}(\ref{eq7}), the '+' sign denotes $in$-$phase$ modulation, and the '-' sign denotes $out$-$of$-$phase$ modulation. $A$ is modulation amplitude and $\omega_m$ represent modulation frequency. For $^{87}$Rb, the scattering lengths are $a_{11}$ = $a_{22}$ = 100.86 $a_0$, $a_{12}$ = $a_{21}$ = 100.4 $a_0$ \cite{bibfech2}, with Bohr radius $a_0$ = 0.0529 nm. The total particle number is $N$ = $10^5$ and we select $ k_r = \sqrt{2\pi}/\lambda$ and $E_r = \hbar k_r^2/2m$ as dimensionless units for length and frequency, respectively. The interaction Hamiltonian is \cite {bib36,bib43}:
	\begin{equation}
		H_{\text{int}} = 
		\begin{pmatrix}
			g_{11}|\psi_1|^2 + g_{12}|\psi_2|^2 & 0 \\
			0 & g_{22}|\psi_2|^2 + g_{12}|\psi_1|^2
		\end{pmatrix}
		.\label{eq2}
	\end{equation}


        \begin{figure}
		\centering
		\includegraphics[width=1\columnwidth]{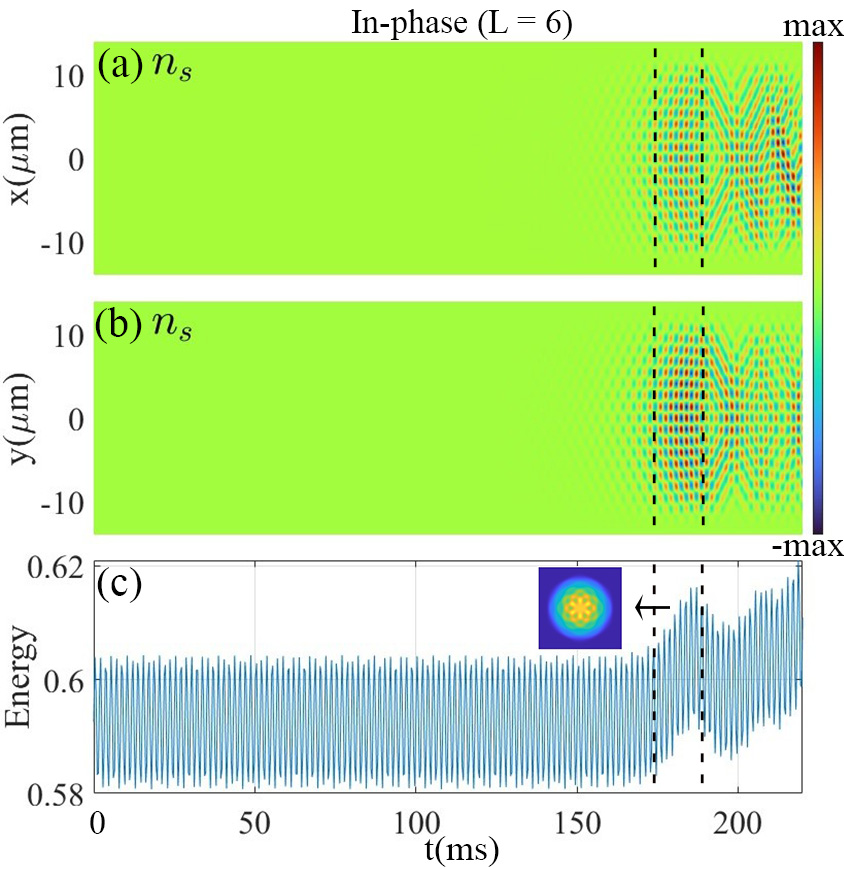}
		\caption{\fontfamily{ptm}\selectfont The time evolution of the system from 0 to 220 ms. The spin Faraday wave is excited with $in$-$phase$ modulation when $\Omega$ = 0.1 $E_r$, $f\equiv \omega _m/2\pi$  = 600 Hz, $A$ = 8$a_0$. (a)(b) The evolution of the density distribution. Periodic variations in the spin density distribution emerge around 174 $\sim$ 190 ms. (c) The evolution of energy over time.}
		\label{fig_revolution}
	\end{figure}

In the SOC BEC system, the ground state energy of a single particle is solved by variational method, distinguishing the three quantum phases: stripe phase, plane wave phase, and zero momentum phase [42]. This paper focuses on the pattern formation of SOC BEC with stripe phase when $\Omega = 0.1 E_r$ and $\delta = 0$. The dispersion relation of the stripe phase in the ground state is shown in \textcolor{blue}{Fig.}\ref{fig_theory}\textcolor{blue}{(b)}. The two lowest energy points correspond to the two spin states. In \textcolor{blue}{Fig.}\ref{fig_theory}\textcolor{blue}{(c)(d)}, the momenta of the two spin states are respectively condensed around $k_x/k_r=\pm1$ in the x-direction.

	The dynamics of the SOC BEC relies on the time-dependent Gross-Pitaevskii (GP) equations:
\begin{equation}
\begin{aligned}
i\hbar\frac{\partial\psi_1}{\partial t} &= \bigg( -\frac{\hbar^2(k_x - k_r)^2}{2m} +\frac{p_{y}^{2} }{2m}+V +\\
&\quad g_{11}|\psi_1|^2  + g_{12}|\psi_2|^2 \bigg) \psi_1 + \frac{\hbar\Omega}{2}\psi_2,
\end{aligned}
\label{eq5}
\end{equation}

\begin{equation}
\begin{aligned}
i\hbar\frac{\partial\psi_2}{\partial t} &= \bigg( -\frac{\hbar^2(k_x + k_r)^2}{2m} +\frac{p_{y}^{2} }{2m}+ V +\\
&\quad g_{22}|\psi_2|^2  + g_{12}|\psi_1|^2 \bigg) \psi_2 + \frac{\hbar\Omega}{2}\psi_1.
\end{aligned}
\label{eq6}
\end{equation}

	\section{Pattern Dynamics}

	\subsection{Faraday wave excitation}
	\label{subsec1}
	 By numerically solving the GP equations, we analyze the excitation of Faraday wave under $in$-$phase$ modulation with $\omega _m/2\pi$ = 600 Hz. \textcolor{blue}{Figure.}\ref{fig_revolution}\textcolor{blue}{(a)(b)} show the evolution of the spin density $n_s = n_1 - n_2$ in the x-direction and y-direction from 0 to 220 ms, where $n_i$ represents $\left | \psi_i  \right | ^2$. When $t < 170\ \text{ms}$, the spin density distributions in the x- and y-directions remain stable, indicating that the system in this stage is insufficient to excite the Faraday pattern. Within 174 $\sim$ 190 ms (black dashed lines), the spin density distributions in the x- and y-directions exhibit periodic variations. However, due to the symmetry breaking of the system caused by the spin-orbit coupling, there are differences in the spin density distributions between the x- and y-directions. \textcolor{blue}{Figure.}\ref{fig_revolution}\textcolor{blue}{(c)} shows the evolution of energy. Before 174 ms, the energy undergoes small-amplitude periodic oscillations. During 174 $\sim$ 190 ms, as the system energy rises, the Faraday pattern with L = 6 is excited within this period. The density distribution of the spin-up component at 179 ms is shown in the illustration of \textcolor{blue}{Fig.}\ref{fig_revolution}\textcolor{blue}{(c)}. After 200 ms, the spin density distribution no longer exhibits periodicity and the energy continues increasing, which indicates that the system has entered the nonlinear regime. 

      \begin{figure}
		\centering
		\includegraphics[width=1\columnwidth]{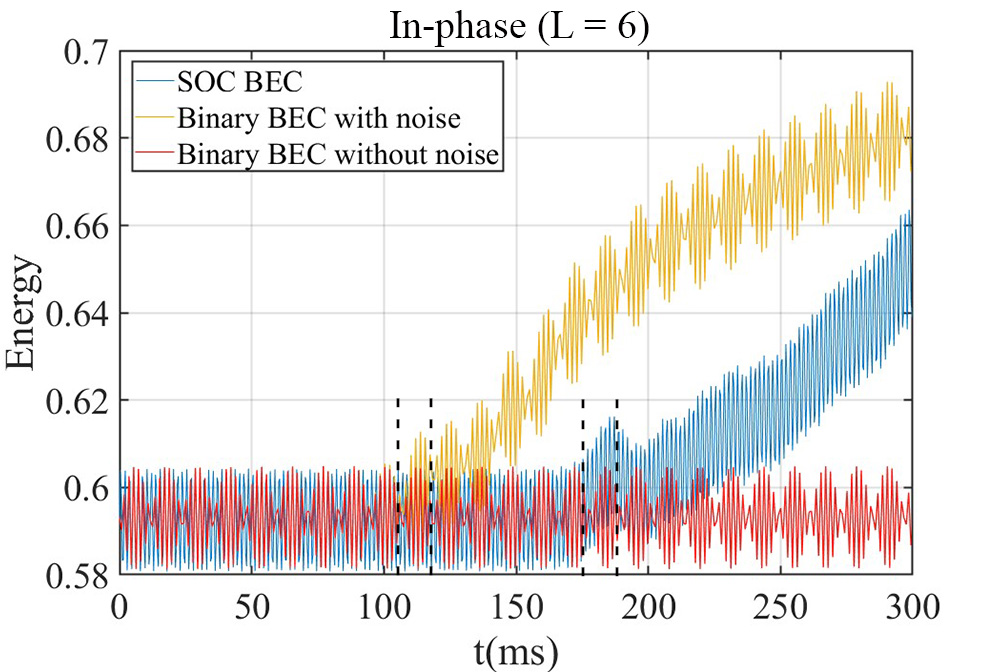}
		\caption{\fontfamily{ptm}\selectfont The energy evolution of different systems when $\omega _m/2\pi$  = 600 Hz, $A$ = 8$a_0$. Blue: $in$-$phase$ modulation without noise (SOC BEC); Yellow: $in$-$phase$ modulation with noise (normal binary BEC); Red: $in$-$phase$ modulation without noise (normal binary BEC).}
		\label{fig_noise}
	\end{figure}

        	\begin{figure*}
		\centering
		\includegraphics[width=1.8\columnwidth]{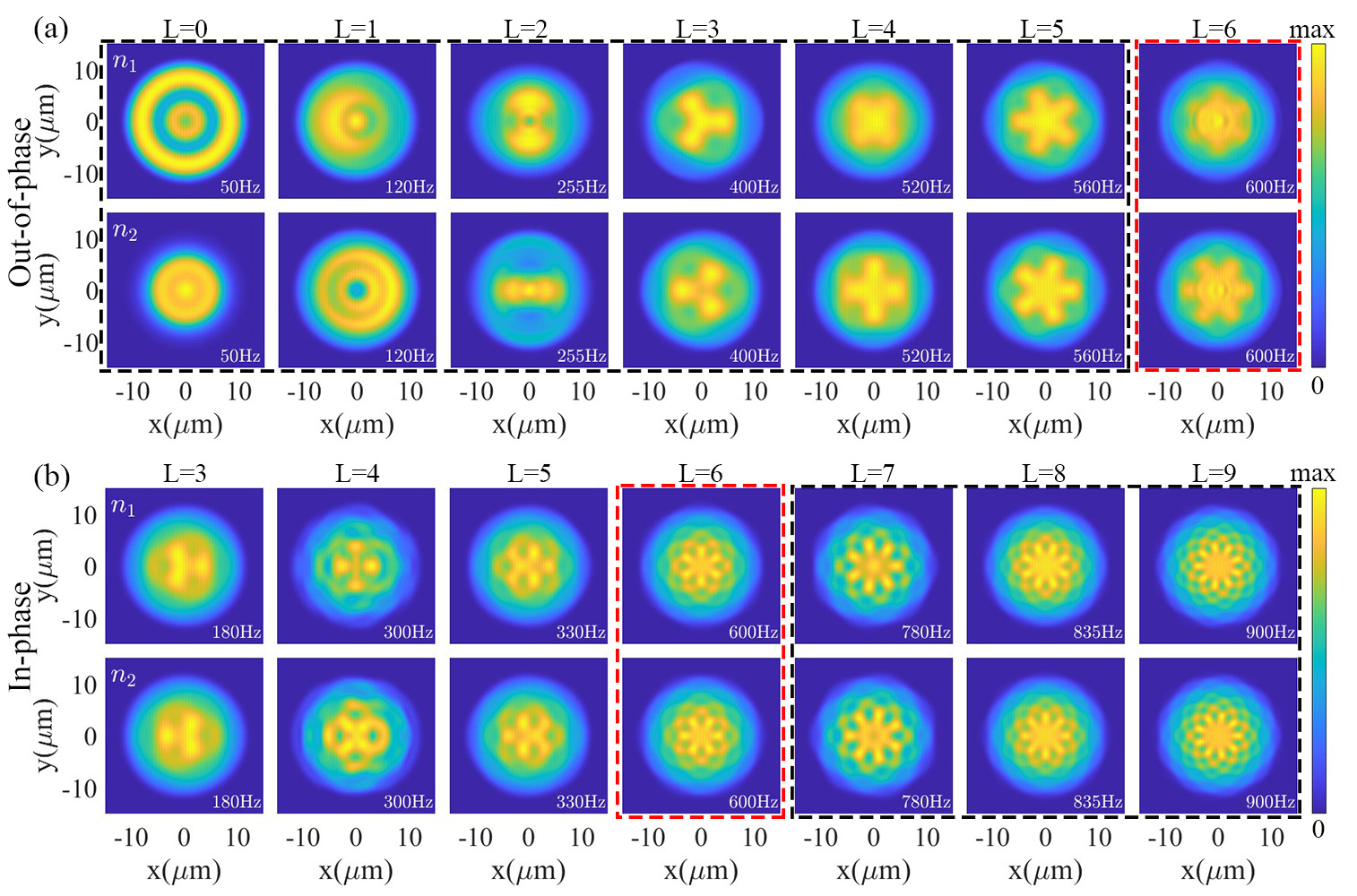}
		\caption{\fontfamily{ptm}\selectfont The density distribution patterns in the row of $n_1$ represent spin-up component, and those in the row of $n_2$ represent spin-down component. Different modulation methods exhibit distinct density distribution patterns and rotational symmetry. (a) $out$-$of$-$phase$ modulation is employed with the corresponding modulation frequencies (from left to right)  $\omega _m/2\pi$  = \{50, 120, 255, 400, 520, 560, 600\} Hz at $t$ = \{16, 553, 284, 390, 414, 370, 304\} ms, where the modulation amplitude is $A$ = 8$a_0$. (b) $in$-$phase$ modulation is employed with the corresponding modulation frequencies (from left to right) $\omega _m/2\pi$  = \{180, 300,330,600,780,835,900\} Hz at $t$ = \{93, 164, 119, 166, 132, 168, 164\} ms, where the modulation amplitude is $A$ = 8$a_0$.}
		\label{fig_pattern}
	\end{figure*}
    Because the generation of patterns requires a slight imbalance of the density distribution, the initial noise is necessary for single-component BEC to produce Faraday waves \cite{bib41.2}. For normal binary BEC, both the phase difference under $out$-$of$-$phase$ modulation and the initial amplitude noise under $in$-$phase$ modulation can disbalance the density distribution and thus to produce spin Faraday patterns \cite{meiling2025}. However, in this paper, due to SOC, the difference of velocities in the x- and y-directions leads to the imbalance of density distribution. $v_x = \frac{\hbar \nabla_x \phi}{m} - \frac{\hbar k_r n_s}{m n_t}$, $ v_y = \frac{\hbar \nabla_y \phi}{m}$ \cite{Qu_2017}, where $n_t=n_1+n_2$ represents the total density. The energy evolutions of different systems with $in$-$phase$ modulation are shown in \textcolor{blue}{Fig.}\ref{fig_noise}. To add some noise, we consider a weak amplitude perturbation to the ground state $\psi_G$. The wave function $\psi_{in}=\psi_G(1+\varepsilon\delta_{in})$. Here $\delta_{in}$ is taken from normally distributed random (with a variance of 1), and $\varepsilon=0.0001$ is the amplitude of the perturbation \cite{bib41.2}. The modulation amplitude affects the speed of pattern excitation in the system, the larger the modulation amplitude, the faster the pattern is excited. We aim to investigate the influence of SOC and noise on pattern excitation, so we fix the modulation amplitude at \(A = 8a_0\). For the normal binary BEC without noise, its energy exhibits periodic oscillations and fails to show an upward trend within 300 ms, indicating that the system fails to produce patterns. The ordinary binary BEC with noise and SOC BEC  successfully generate the Faraday patterns at 100 $\sim$ 116 ms and 174 $\sim$ 190 ms, respectively, and the energy continues to rise. Compared with the ordinary binary BEC system, both SOC and noise break the symmetry of system, making it easier to generate the polygonal patterns.

Next, the patterns excited via two modulation schemes are presented in \textcolor{blue}{Fig.}\ref{fig_pattern}, all the patterns are symmetric about the x-axis in SOC systems. Similar to the time period within the black dashed lines in \textcolor{blue}{Fig.}\ref{fig_revolution}, all the patterns are selected during the period when the system has been generated but not yet entered the nonlinear destabilization regime. Through $out$-$of$-$phase$ modulation and $in$-$phase$ modulation, we respectively select the patterns with different rotational symmetries from L = 0 to L = 6 and from L = 3 to L = 9. It should be noted that the system exhibits subharmonicity. Under $in$-$phase$ modulation the natural frequencies of the patterns in \textcolor{blue}{Fig.}\ref{fig_pattern}\textcolor{blue}{(b)} are
$\omega_0/2\pi=1/T$ = \{95, 148, 164, 309, 394, 412, 439\} Hz, satisfying the Faraday wave resonance relation $\omega_m\simeq2\omega_0$. Whereas under $out$-$of$-$phase$ modulation, the resonant patterns will emerge as shown in \textcolor{blue}{Fig.}\ref{fig_pattern}\textcolor{blue}{(a)} with $\omega_0/2\pi=1/T$ = \{56, 119, 265, 403, 531, 568, 613\} Hz, satisfying the resonant wave relation $\omega_m\simeq\omega_0$ \cite{PhysRevE.84.056202}.

\

    \subsection{The symmetry of patterns}
	\label{subsec2}

          	\begin{figure}
		\centering
		\includegraphics[width=0.95\columnwidth]{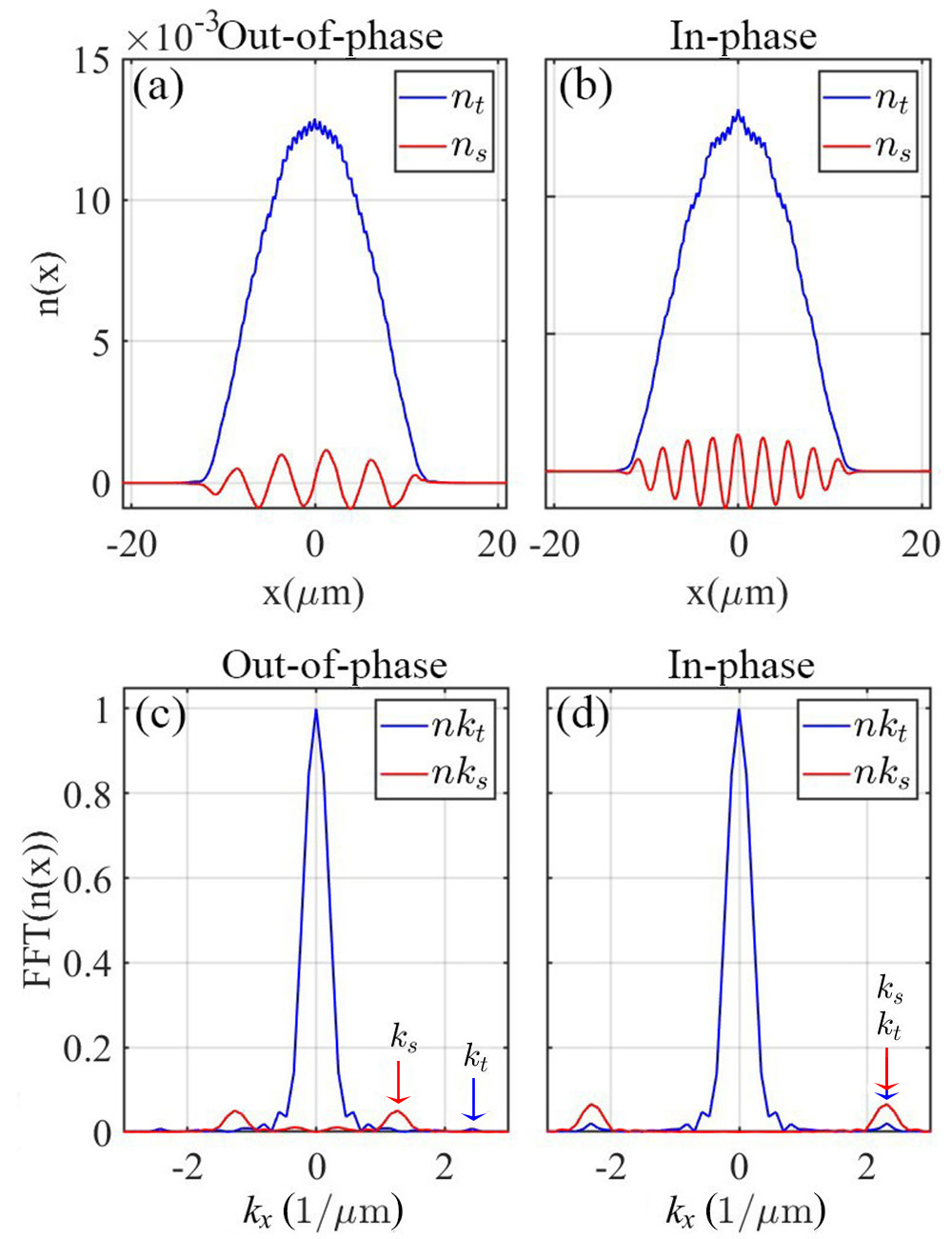}
		\caption{\fontfamily{ptm}\selectfont A comparison of the $out$-$of$-$phase$ and $in$-$phase$ modulation with L = 6 and $\omega _m/2\pi$  = 600 Hz. (a)(b) Total density and spin density in the x-direction, $n_t(x)=\int (n_1+n_2)dy$, $n_s(x)=\int (n_1-n_2)dy$. (c)(d) Fourier transform of the $n_t$ and $n_s$.}
		\label{fig_1D}
	\end{figure}
	Owing to strong dissipation, the patterns with L = 6 are difficult to produce in classical fluid \cite{bib41.1}. In our system, the same phenomenon is revealed. As shown by the red dashed line in \textcolor{blue}{Fig.}\ref{fig_pattern}\textcolor{blue}{(a)}, the symmetry of the pattern with L = 6 is broken under $out$-$of$-$phase$ modulation. Moreover, by analyzing a large amount of data with $f\equiv \omega _m/2\pi$ ranging from 0 - 1000 Hz, we find that the patterns with L $>$ 6 cannot be successfully generated. Notably, the patterns with higher-order rotational symmetries L = \{6, 7, 8, 9\} are excited under $in$-$phase$ modulation, and patterns with L $\geq$ 6 exhibit high symmetry in \textcolor{blue}{Fig.}\ref{fig_pattern}\textcolor{blue}{(b)}. Thus, L = 6 is a special critical value.

\begin{figure}
		\centering
		\includegraphics[width=0.9\columnwidth]{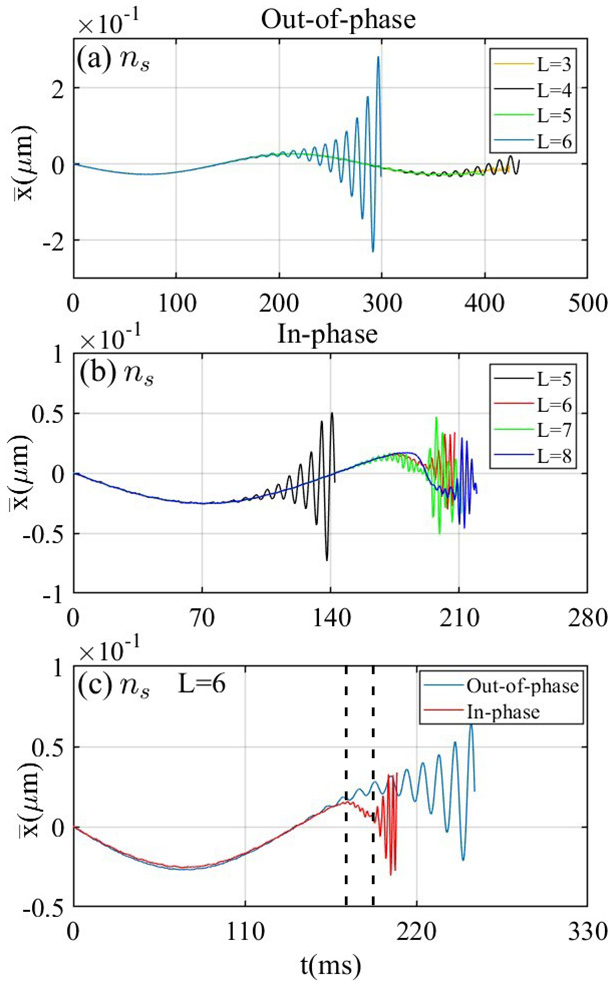}
		\caption{\fontfamily{ptm}\selectfont (a) Mass center evolution of spin density \( n_s \) along the x-direction under $out$-$of$-$phase$ modulation. For L = 6, the system exhibits a strong dipole mode, causing the symmetry breaking of the pattern.  
(b) Mass center evolution of spin density $n_s$ along the x-direction under in-phase modulation.  
(c) Comparison of mass center evolutions between $in$-$phase$ and $out$-$of$-$phase$ modulations for L = 6.}
		\label{fig_mass}
	\end{figure}
    \textcolor{blue}{Figure.}\ref{fig_1D} shows the x-direction characteristics of the patterns with L = 6 under different modulation schemes. The total density $n_t$ of the two components exhibit a striped distribution along the x-direction in \textcolor{blue}{Fig.}\ref{fig_1D}\textcolor{blue}{(a)(b)}, which represents the breaking of continuous translational symmetry. While the breaking of continuous translational symmetry is a typical feature of supersolids \cite{bib36.5}. In our SOC system with stripe phase, all the patterns with different rotational symmetries exhibit the characteristics of supersolids. 

    We perform Fourier transforms on the spin density $n_s$ and total density $n_t$ to obtain the momentum distribution of spin and density waves. As shown in \textcolor{blue}{Fig.}\ref{fig_1D}\textcolor{blue}{(c)(d)}, the spin and density waves propagate radially, with $k_s$ and $k_t$ denoting the momenta of the spin and density waves, respectively. The red arrow $k_s$ and blue arrow $k_t$ denote the two wave peaks of the spin wave and density wave, respectively. Under $out$-$of$-$phase$ modulation, $k_s = 1.2729\ \mu\mathrm{m}^{-1}$ and $k_t = 2.5430\ \mu\mathrm{m}^{-1}$, indicating that the wave vector of the spin wave is approximately half that of the density wave $2k_s \simeq k_t$. In our numerical simulations, for L = 6 with $in$-$phase$ modulation, the difference of wave vector $\Delta k = k_t - k_s = 0$. 

    The symmetry breaking at L = 6 has been studied in classical fluid \cite{bib41.1} and single-component BEC \cite{bib41.2}. Through $\bar{x}_{n_i}=\int(xn_i)d\mathbf{r}$, $\bar{y}_{n_i}=\int(yn_i)d\mathbf{r}$, we obtain the evolution relationship of the mass center offset in the x- and y-directions over time. Since the SOC occurs in the x-direction, the offset of the mass center in the y-direction we obtained is extremely small, oscillating at the order of \(10^{-9}\) $\mu m$, which can be neglected.

    As shown in \textcolor{blue}{Fig.}\ref{fig_mass}, due to SOC, the mass centers of the two components periodically exchange along the x-direction, leading to the oscillation of the mass center of $n_s$ with a period $ T_m \approx 280$ ms (the curve of the system entering the nonlinear destabilization regime is not plotted in the figure). We calculate the mass center evolution of the system when the rotational symmetry changes from L = 3 to L = 6 under $out$-$of$-$phase$ modulation. In \textcolor{blue}{Fig.}\ref{fig_mass}\textcolor{blue}{(a)}, the oscillation of the mass center is relatively weak when L = \{3, 4, 5\}. However, when L = 6, an obvious dipolar mode appears after 200 ms, causing the mass center of the system to deviate significantly. This is the reason for the symmetry breaking in \textcolor{blue}{Fig.}\ref{fig_pattern}\textcolor{blue}{(a)} when L = 6.

    Similarly, as shown in \textcolor{blue}{Fig.}\ref{fig_mass}\textcolor{blue}{(b)}, the pattern is excited around 100 ms when L = 5. The dipole oscillation also appears simultaneously and its amplitude gradually increases, leading to the symmetry breaking in \textcolor{blue}{Fig.}\ref{fig_pattern}\textcolor{blue}{(b)}. However, the pattern and dipole oscillation emerge successively when L $\ge$ 6 with $in$-$phase$ modulation. \textcolor{blue}{Figure.}\ref{fig_mass}\textcolor{blue}{(c)} shows the comparison of mass center evolutions between $in$-$phase$ (red curve) and $out$-$of$-$phase$ (blue curve) modulations for L = 6. During 174 - 190 ms (black dashed line), the mass center of the system exhibits only small-amplitude oscillations with $in$-$phase$ modulation, corresponding to the excitation of the Faraday pattern in \textcolor{blue}{Fig.}\ref{fig_revolution}. The dipole mode suddenly emerges after 190 ms, causing the system to become unstable. The excitation of the Faraday pattern precedes the emergence of the dipole mode, which makes the pattern highly symmetric. Notably, with the same parameters, $in$-$phase$ modulation generates patterns more rapidly than $out$-$of$-$phase$ modulation.

		\subsection{The pattern nodes and radius}
	\label{subsec3}
	 Freezing the rotational symmetry L while varying the number of nodes is a characteristic in the single-component BEC \cite{node1} and binary BEC \cite{meiling2025}. Building on this framework, we propose that in our SOC BEC system, a similar approach of freezing L can be employed to explore node-dependent excitation modes. Under $out$-$of$-$phase$ modulation, the radial node number $n_r$ varies exclusively within the range of 1 to 2, and the system cannot generate surface modes with more number of nodes. However, under $in$-$phase$ modulation, we successfully generate high-order rotational symmetry patterns with L $\ge$ 6, and the mode of nodes is more diverse when L = 9.

          \begin{figure}
		\centering
		\includegraphics[width=1\columnwidth]{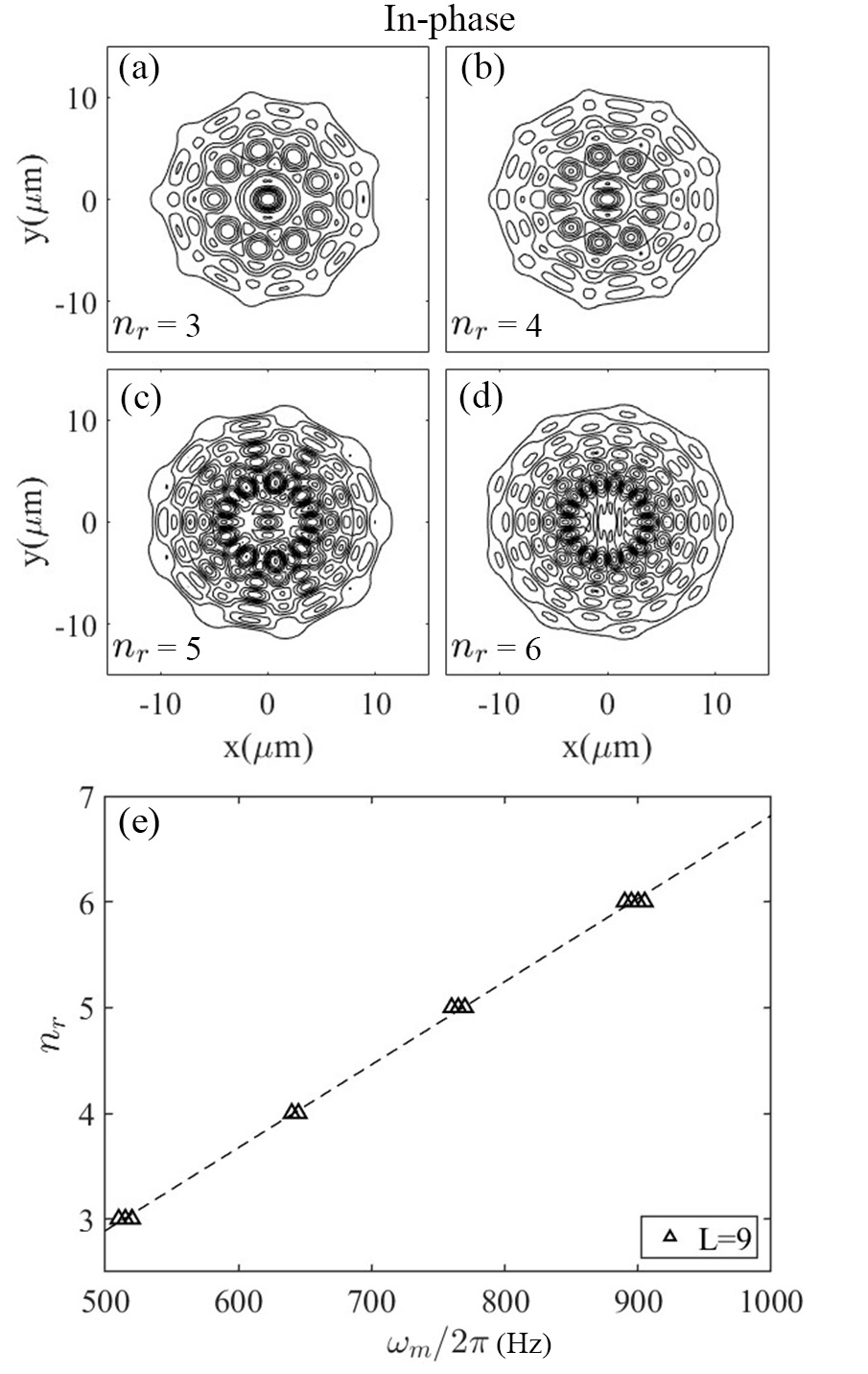}
		\caption{\fontfamily{ptm}\selectfont (a)(b)(c)(d) shows the equipotential line of density distribution with $n_r$ = \{3, 4, 5, 6\} under $in$-$phase$ modulation when L = 9. (e) The relation of the radial nodes $n_r$  with the modulation frequency $\omega _m/2\pi$  from 500 to 1000 Hz under $in$-$phase$ modulation when L = 9.}
		\label{fig_nodes}
	\end{figure}
    
		\begin{figure}
		\centering
		\includegraphics[width=1\columnwidth]{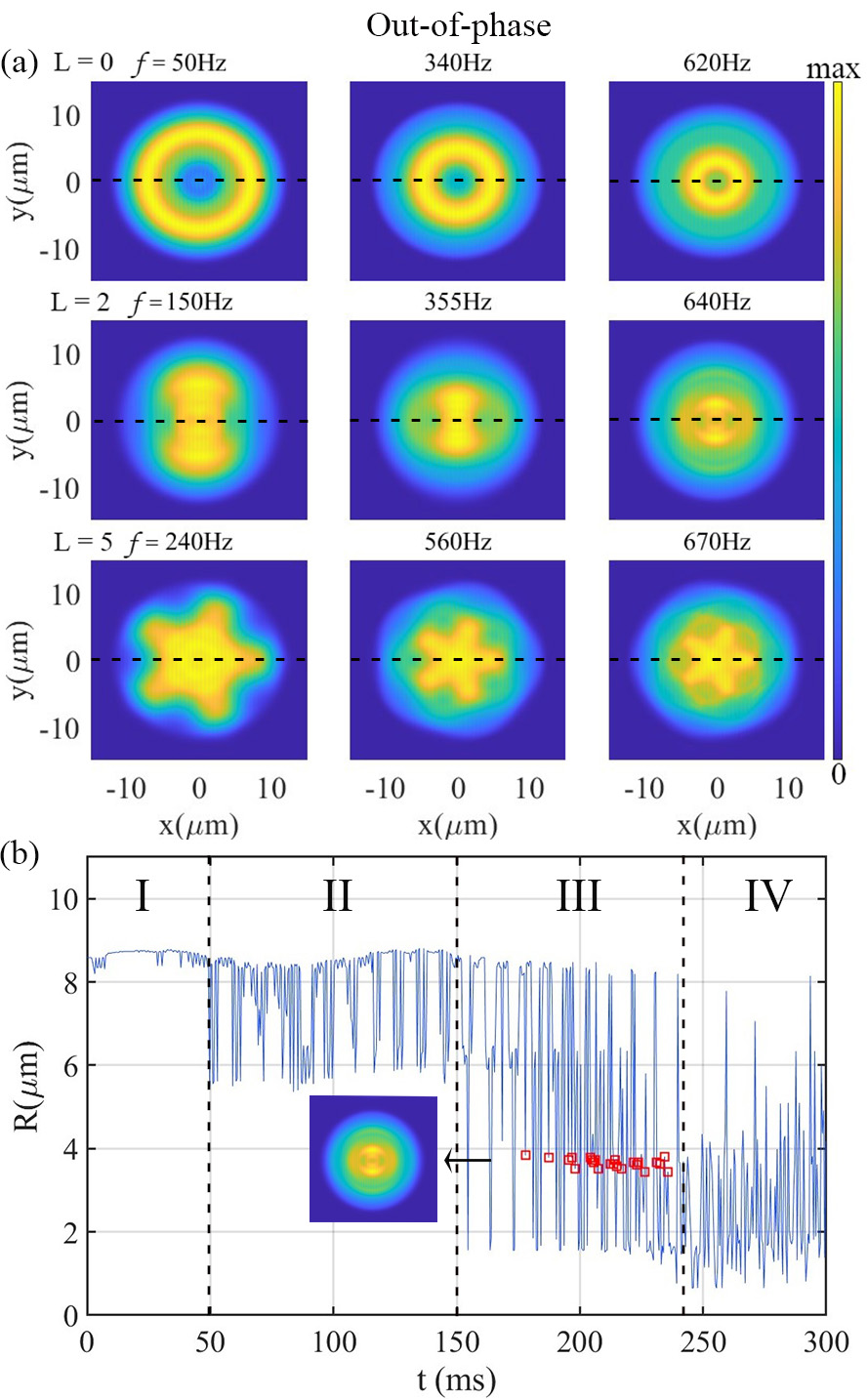}
		\caption{\fontfamily{ptm}\selectfont (a) Pattern radii vary with $\omega _m/2\pi$  = \{50, 340, 620, 150, 355, 640, 240, 560, 670\} Hz under $out$-$of$-$phase$ modulation. The rotational symmetry of each row is L = \{0, 2, 5\}. The modulation frequency is inversely proportional to the pattern radius with the same L. (b) The pattern radius varies with time under $out$-$of$-$phase$ modulation with $\omega _m/2\pi$ = 640 Hz. \textcolor{blue}{I} the non-excited state, \textcolor{blue}{II} breathing modes of varying radius, \textcolor{blue}{III} the emergence of patterns, \textcolor{blue}{IV} the nonlinear destabilization regime.}
		\label{fig_radius}
	\end{figure}
   By plotting the equipotential lines of the density distribution, we can clearly visualize the number of nodes. \textcolor{blue}{Figure.}\ref{fig_nodes}\textcolor{blue}{(a)(b)(c)(d)} show the number of nodes $n_r$ = \{3, 4, 5, 6\} respectively when L = 9 with modulation frequency $\omega_m/2\pi$ = \{520, 640, 760, 900\}. The relation between the number of nodes $n_r$ and modulation frequency $\omega_m/2\pi$ is shown by the scatter points in \textcolor{blue}{Fig.}\ref{fig_nodes}\textcolor{blue}{(e)} when L = 9. The least squares fitting reveals a high consistency between the scatter points and a linear model: $n_r = \tau \omega_m + \eta$, where $\tau $ denotes the slope and $\eta$ denotes a constant. When L = 9, the parameters are $\tau =7.85\times 10^{-3}$, and $\eta=-1.03$. The slope $\tau >0$ indicates that, for a fixed rotational symmetry L, the number of nodes $n_r$ increases with the modulation frequency $\omega_m$.

    Modifying the trapping potential represents an effective approach to control the radius of BEC patterns \cite{bib48,bib49}. However, we find that under $out$-$of$-$phase$ modulation, the pattern radius can be controlled by changing the modulation frequency  \(\omega_m\). \textcolor{blue}{Fig.}\ref{fig_radius}\textcolor{blue}{(a)} exhibits patterns with the same rotational symmetry L but distinct radii.

   To calculate the radius of a polygonal pattern, we select points satisfying \(\delta(x,y) = \nabla_{x,y}|\psi(x,y)|^2\) to obtain $n$ points \((x_i, y_i)\) with the largest density gradient changes \cite{9114}. The distance from each point to the pattern center is calculated as \(R_i = \sqrt{x_i^2 + y_i^2}\), and the maximum value of \(R_i\) is taken as the pattern radius. The radius evolution is shown in \textcolor{blue}{Fig.}\ref{fig_radius}\textcolor{blue}{(b)}. Between 160 $\sim$ 250 ms, patterns appear alongside three intensities of breathing modes. This includes a weak breathing mode with a radius $R_I\simeq 8.3$ $\mu$m in stage \textcolor{blue}{I}, a breathing mode with $R_{II}\simeq 5.6$ $\mu$m in stage \textcolor{blue}{II}, and a strong breathing mode with $R_{III}\simeq 1.8$ $\mu$m in stage \textcolor{blue}{III}. All the radii of patterns in stage \textcolor{blue}{III} are shown by the red triangular scatter points in \textcolor{blue}{Fig.}\ref{fig_radius}\textcolor{blue}{(b)}. After 250 ms, the pattern starts to become unstable and enters the nonlinear destabilization regime. When rotational symmetry L is fixed, we calculate the radius of each row in \textcolor{blue}{Fig.}\ref{fig_radius}\textcolor{blue}{(a)}. $R_{L=0}=\{8.89, 6.12, 4.17\}$ $\mu$m, $R_{L=2}=\{8.94, 5.47, 3.72\}$ $\mu$m, $R_{L=5}=\{9.89, 6.93, 5.81\}$ $\mu$m. It can be concluded that when the patterns exhibit the same rotational symmetry L, there is a negative correlation between the modulation frequency and the excitation radius.

	\section{Conclusion}
	In this paper, we numerically studied the pattern formation of spin Faraday waves in a periodically driven pancake-shaped SOC BEC in stripe phase with modulation frequencies range from 0 to 1000 Hz. We analyze the pattern dynamics when the atomic interaction is modulated using two different schemes. The collective excitation modes exhibit different L-fold rotational symmetries at certain frequencies.

We find L = 6 is a special critical value for $in$-$phase$ and $out$-$of$-$phase$ modulations with $\omega _m/2\pi$  = 600 Hz and $A$ = 8$a_0$.  Under $out$-$of$-$phase$ modulation, the rotational symmetry patterns are excited when L $<$ 6, but the symmetry of the pattern with L = 6 loses its symmetry and becomes unstable.  The same situation occurs in the single-component BEC due to the appearance of dipole mode \cite{bib41.2}. But in SOC BEC with $in$-$phase$ modulation, Faraday mode and dipole mode emerge individually in different period, leading to the highly symmetric patterns with L = 6.

In addition, under $in$-$phase$ modulation, we can generate Faraday patterns when L = \{7, 8, 9\} with higher-order rotational symmetry, which are difficult to obtain in previous studies \cite{bib41.1,bib41.2}. Moreover, different from normal binary BEC, Faraday patterns with stripe phase of SOC systems can be excited under $in$-$phase$ modulation without the initial noise \cite{meiling2025}.
Noted that with $in$-$phase$ modulation at certain L, the number of radial nodes $n_r$ increases with the modulation frequency $\omega_m$ increasing.
And with $out$-$of$-$phase$ modulation at certain L, the modulation frequency is inversely proportional to the pattern radius. Different from the way of modulating the potential \cite{bib48,bib49}, we manipulate the radius of patterns by periodically modulating the atomic interaction.

In this work, spin Faraday patterns are successfully excited by applying periodic driving to a SOC BEC confined in a harmonic potential trap. These patterns exhibit supersolid characteristics. We find that tuning the atomic interactions can significantly alter key features of the patterns, including their symmetry, number of nodes, and radius, thereby providing an ideal and versatile platform for exploring supersolid patterns. Furthermore, the dynamics of Faraday patterns under different coupling strengths, as well as the influence of external potentials and dissipation effects on pattern stability, constitute key future research directions.

	\section*{Acknowledgement}
 S.C., H.L. are joint first authors with equal contributions. Our work is supported by the National Key Research and Development Program of China (Grant No. 2025YFF0515201), the Joint Funds of the National Natural Science Foundation of China (Grant No. U25D8014), the National Natural Science Foundation of China (Grant No. 11774093), the Natural Science Foundation of Shanghai (Grant No. 23ZR1418700), the Program of Chongqing Natural Science Foundation (Grant No. CSTB2022NSCQ-MSX0585).

	\bibliographystyle{model1-num-names}
	
	\bibliography{cas-refs}

\begin{thebibliography}{55}
\expandafter\ifx\csname natexlab\endcsname\relax\def\natexlab#1{#1}\fi
\providecommand{\url}[1]{\texttt{#1}}
\providecommand{\href}[2]{#2}
\providecommand{\path}[1]{#1}
\providecommand{\DOIprefix}{doi:}
\providecommand{\ArXivprefix}{arXiv:}
\providecommand{\URLprefix}{URL: }
\providecommand{\Pubmedprefix}{pmid:}
\providecommand{\doi}[1]{\href{http://dx.doi.org/#1}{\path{#1}}}
\providecommand{\Pubmed}[1]{\href{pmid:#1}{\path{#1}}}
\providecommand{\bibinfo}[2]{#2}
\ifx\xfnm\relax \def\xfnm[#1]{\unskip,\space#1}\fi
\bibitem[{Ji and Li(2004)}]{bib10}
\bibinfo{author}{L.~Ji}, \bibinfo{author}{Q.~Li},
\newblock \bibinfo{title}{Effect of local feedback on {Turing} pattern
  formation},
\newblock \bibinfo{journal}{Chem. Phys. Lett.} \bibinfo{volume}{391}
  (\bibinfo{year}{2004}) \bibinfo{pages}{176--180}.
\bibitem[{Sornborger and Parry(1999)}]{bib7}
\bibinfo{author}{A.~Sornborger}, \bibinfo{author}{M.~Parry},
\newblock \bibinfo{title}{Patterns from preheating},
\newblock \bibinfo{journal}{Phys. Rev. Lett.} \bibinfo{volume}{83}
  (\bibinfo{year}{1999}) \bibinfo{pages}{666--669}.
\bibitem[{Carr et~al.(2018)Carr, Massatt, Torrisi, Cazeaux, Luskin, and
  Kaxiras}]{bib4}
\bibinfo{author}{S.~Carr}, \bibinfo{author}{D.~Massatt}, \bibinfo{author}{S.~B.
  Torrisi}, \bibinfo{author}{P.~Cazeaux}, \bibinfo{author}{M.~Luskin},
  \bibinfo{author}{E.~Kaxiras},
\newblock \bibinfo{title}{Relaxation and domain formation in incommensurate
  two-dimensional heterostructures},
\newblock \bibinfo{journal}{Phys. Rev. B} \bibinfo{volume}{98}
  (\bibinfo{year}{2018}) \bibinfo{pages}{224102}.
\bibitem[{Faraday(1831)}]{bib11}
\bibinfo{author}{M.~Faraday},
\newblock \bibinfo{title}{On a peculiar class of acoustical figures; and on
  certain forms assumed by groups of particles upon vibrating elastic
  surfaces},
\newblock \bibinfo{journal}{Philos. Trans. R. Soc. Lond} \bibinfo{volume}{121}
  (\bibinfo{year}{1831}) \bibinfo{pages}{299--340}.
\bibitem[{Cross and Hohenberg(1993)}]{bib1}
\bibinfo{author}{M.~C. Cross}, \bibinfo{author}{P.~C. Hohenberg},
\newblock \bibinfo{title}{Pattern formation outside of equilibrium},
\newblock \bibinfo{journal}{Rev. Mod. Phys.} \bibinfo{volume}{65}
  (\bibinfo{year}{1993}) \bibinfo{pages}{851--1112}.
\bibitem[{Zhang et~al.(2020)Zhang, Yao, Feng, Hu, and Chin}]{bib2}
\bibinfo{author}{Z.~Zhang}, \bibinfo{author}{K.-X. Yao},
  \bibinfo{author}{L.~Feng}, \bibinfo{author}{J.~Hu},
  \bibinfo{author}{C.~Chin},
\newblock \bibinfo{title}{Pattern formation in a driven {Bose}-{Einstein}
  condensate},
\newblock \bibinfo{journal}{Nat. Phys.} \bibinfo{volume}{16}
  (\bibinfo{year}{2020}) \bibinfo{pages}{652--655}.
\bibitem[{Westra et~al.(2003)Westra, Binks, and Van De~Water}]{bib8}
\bibinfo{author}{M.~Westra}, \bibinfo{author}{D.~J. Binks},
  \bibinfo{author}{W.~Van De~Water},
\newblock \bibinfo{title}{Patterns of {Faraday} waves},
\newblock \bibinfo{journal}{J. Fluid Mech} \bibinfo{volume}{496}
  (\bibinfo{year}{2003}) \bibinfo{pages}{1–32}.
\bibitem[{Benjamin et~al.(1954)Benjamin, Ursell, and Taylor}]{bib12}
\bibinfo{author}{T.~B. Benjamin}, \bibinfo{author}{F.~J. Ursell},
  \bibinfo{author}{G.~I. Taylor},
\newblock \bibinfo{title}{The stability of the plane free surface of a liquid
  in vertical periodic motion},
\newblock \bibinfo{journal}{Proc. R. Soc. Lond. A} \bibinfo{volume}{225}
  (\bibinfo{year}{1954}) \bibinfo{pages}{505--515}.
\bibitem[{Anderson et~al.(1995)Anderson, Ensher, Matthews, Wieman, and
  Cornell}]{bib13}
\bibinfo{author}{M.~H. Anderson}, \bibinfo{author}{J.~R. Ensher},
  \bibinfo{author}{M.~R. Matthews}, \bibinfo{author}{C.~E. Wieman},
  \bibinfo{author}{E.~A. Cornell},
\newblock \bibinfo{title}{Observation of {Bose}-{Einstein} condensation in a
  dilute atomic vapor},
\newblock \bibinfo{journal}{Science} \bibinfo{volume}{269}
  (\bibinfo{year}{1995}) \bibinfo{pages}{198--201}.
\bibitem[{Kartashov and Konotop(2017)}]{bib14}
\bibinfo{author}{Y.~V. Kartashov}, \bibinfo{author}{V.~V. Konotop},
\newblock \bibinfo{title}{Solitons in {Bose}-{Einstein} condensates with
  helicoidal spin-orbit coupling},
\newblock \bibinfo{journal}{Phys. Rev. Lett.} \bibinfo{volume}{118}
  (\bibinfo{year}{2017}) \bibinfo{pages}{190401}.
\bibitem[{Weiler et~al.(2008)Weiler, Neely, Scherer, Bradley, Davis, and
  Anderson}]{bib15}
\bibinfo{author}{C.~N. Weiler}, \bibinfo{author}{T.~W. Neely},
  \bibinfo{author}{D.~R. Scherer}, \bibinfo{author}{A.~S. Bradley},
  \bibinfo{author}{M.~J. Davis}, \bibinfo{author}{B.~P. Anderson},
\newblock \bibinfo{title}{Spontaneous vortices in the formation of
  {Bose}–{Einstein} condensates},
\newblock \bibinfo{journal}{Nature} \bibinfo{volume}{455}
  (\bibinfo{year}{2008}) \bibinfo{pages}{948--951}.
\bibitem[{Muryshev et~al.(2002)Muryshev, Shlyapnikov, Ertmer, Sengstock, and
  Lewenstein}]{bib16}
\bibinfo{author}{A.~Muryshev}, \bibinfo{author}{G.~V. Shlyapnikov},
  \bibinfo{author}{W.~Ertmer}, \bibinfo{author}{K.~Sengstock},
  \bibinfo{author}{M.~Lewenstein},
\newblock \bibinfo{title}{Dynamics of dark solitons in elongated
  {Bose}-{Einstein} condensates},
\newblock \bibinfo{journal}{Phys. Rev. Lett.} \bibinfo{volume}{89}
  (\bibinfo{year}{2002}) \bibinfo{pages}{110401}.
\bibitem[{Raman et~al.(2001)Raman, Abo-Shaeer, Vogels, Xu, and
  Ketterle}]{bib17}
\bibinfo{author}{C.~Raman}, \bibinfo{author}{J.~R. Abo-Shaeer},
  \bibinfo{author}{J.~M. Vogels}, \bibinfo{author}{K.~Xu},
  \bibinfo{author}{W.~Ketterle},
\newblock \bibinfo{title}{Vortex nucleation in a stirred {Bose}-{Einstein}
  condensate},
\newblock \bibinfo{journal}{Phys. Rev. Lett.} \bibinfo{volume}{87}
  (\bibinfo{year}{2001}) \bibinfo{pages}{210402}.
\bibitem[{Li et~al.(2021)Li, Deng, Zhang, Wu, Xia, and Yi}]{wuhaibin}
\bibinfo{author}{F.~Li}, \bibinfo{author}{S.~Deng}, \bibinfo{author}{L.~Zhang},
  \bibinfo{author}{H.~Wu}, \bibinfo{author}{J.~Xia}, \bibinfo{author}{L.~Yi},
\newblock \bibinfo{title}{Light induced space-time patterns in a superfluid
  {Fermi} gas},
\newblock \bibinfo{journal}{Sci. China-Phys. Mech. Astron.}
  \bibinfo{volume}{64} (\bibinfo{year}{2021}) \bibinfo{pages}{294212}.
\bibitem[{Li et~al.(2025)Li, Wen, Wan, Li, and Wang}]{wenwen}
\bibinfo{author}{H.~Li}, \bibinfo{author}{W.~Wen}, \bibinfo{author}{J.~Wan},
  \bibinfo{author}{H.~Li}, \bibinfo{author}{Y.~Wang},
\newblock \bibinfo{title}{Stability diagram and parametric excitation in a
  strongly interacting fermionic superfluid},
\newblock \bibinfo{journal}{Phys. A: Stat. Mech. Appl.} \bibinfo{volume}{669}
  (\bibinfo{year}{2025}) \bibinfo{pages}{130590}.
\bibitem[{Staliunas et~al.(2002)Staliunas, Longhi, and de~Valc\'arcel}]{bib19}
\bibinfo{author}{K.~Staliunas}, \bibinfo{author}{S.~Longhi},
  \bibinfo{author}{G.~J. de~Valc\'arcel},
\newblock \bibinfo{title}{Faraday patterns in {Bose}-{Einstein} condensates},
\newblock \bibinfo{journal}{Phys. Rev. Lett.} \bibinfo{volume}{89}
  (\bibinfo{year}{2002}) \bibinfo{pages}{210406}.
\bibitem[{Douady(1990)}]{bib9}
\bibinfo{author}{S.~Douady},
\newblock \bibinfo{title}{Experimental study of the {Faraday} instability},
\newblock \bibinfo{journal}{J. Fluid Mech} \bibinfo{volume}{221}
  (\bibinfo{year}{1990}) \bibinfo{pages}{383 -- 409}.
\bibitem[{Engels et~al.(2007)Engels, Atherton, and Hoefer}]{bib20}
\bibinfo{author}{P.~Engels}, \bibinfo{author}{C.~Atherton},
  \bibinfo{author}{M.~A. Hoefer},
\newblock \bibinfo{title}{Observation of {Faraday} waves in a {Bose}-{Einstein}
  condensate},
\newblock \bibinfo{journal}{Phys. Rev. Lett.} \bibinfo{volume}{98}
  (\bibinfo{year}{2007}) \bibinfo{pages}{095301}.
\bibitem[{Nicolin et~al.(2007)Nicolin, Carretero-Gonz\'alez, and
  Kevrekidis}]{bib20.1}
\bibinfo{author}{A.~I. Nicolin}, \bibinfo{author}{R.~Carretero-Gonz\'alez},
  \bibinfo{author}{P.~G. Kevrekidis},
\newblock \bibinfo{title}{Faraday waves in {Bose}-{Einstein} condensates},
\newblock \bibinfo{journal}{Phys. Rev. A} \bibinfo{volume}{76}
  (\bibinfo{year}{2007}) \bibinfo{pages}{063609}.
\bibitem[{Recati and Stringari(2022)}]{bib21}
\bibinfo{author}{A.~Recati}, \bibinfo{author}{S.~Stringari},
\newblock \bibinfo{title}{Coherently coupled mixtures of ultracold atomic
  gases},
\newblock \bibinfo{journal}{Annu. Rev. Condens. Matter Phys.}
  \bibinfo{volume}{13} (\bibinfo{year}{2022}) \bibinfo{pages}{407--432}.
\bibitem[{Cominotti et~al.(2022)Cominotti, Berti, Farolfi, Zenesini, Lamporesi,
  Carusotto, Recati, and Ferrari}]{bib23}
\bibinfo{author}{R.~Cominotti}, \bibinfo{author}{A.~Berti},
  \bibinfo{author}{A.~Farolfi}, \bibinfo{author}{A.~Zenesini},
  \bibinfo{author}{G.~Lamporesi}, \bibinfo{author}{I.~Carusotto},
  \bibinfo{author}{A.~Recati}, \bibinfo{author}{G.~Ferrari},
\newblock \bibinfo{title}{Observation of massless and massive collective
  excitations with {Faraday} patterns in a two-component superfluid},
\newblock \bibinfo{journal}{Phys. Rev. Lett.} \bibinfo{volume}{128}
  (\bibinfo{year}{2022}) \bibinfo{pages}{210401}.
\bibitem[{Maity et~al.(2020)Maity, Mukherjee, Mistakidis, Das, Kevrekidis,
  Majumder, and Schmelcher}]{bib44}
\bibinfo{author}{D.~K. Maity}, \bibinfo{author}{K.~Mukherjee},
  \bibinfo{author}{S.~I. Mistakidis}, \bibinfo{author}{S.~Das},
  \bibinfo{author}{P.~G. Kevrekidis}, \bibinfo{author}{S.~Majumder},
  \bibinfo{author}{P.~Schmelcher},
\newblock \bibinfo{title}{Parametrically excited star-shaped patterns at the
  interface of binary {Bose}-{Einstein} condensates},
\newblock \bibinfo{journal}{Phys. Rev. A} \bibinfo{volume}{102}
  (\bibinfo{year}{2020}) \bibinfo{pages}{033320}.
\bibitem[{Lin et~al.(2009)Lin, Compton, Jimenez-Garcia, Porto, and
  Spielman}]{bib24}
\bibinfo{author}{Y.~J. Lin}, \bibinfo{author}{R.~L. Compton},
  \bibinfo{author}{K.~Jimenez-Garcia}, \bibinfo{author}{J.~V. Porto},
  \bibinfo{author}{I.~B. Spielman},
\newblock \bibinfo{title}{Synthetic magnetic fields for ultracold neutral
  atoms},
\newblock \bibinfo{journal}{Nature} \bibinfo{volume}{462}
  (\bibinfo{year}{2009}) \bibinfo{pages}{628--632}.
\bibitem[{Lin et~al.(2011)Lin, Jimenez-Garcia, and Spielman}]{bib25}
\bibinfo{author}{Y.~J. Lin}, \bibinfo{author}{K.~Jimenez-Garcia},
  \bibinfo{author}{I.~B. Spielman},
\newblock \bibinfo{title}{Spin-orbit-coupled {Bose}-{Einstein} condensates},
\newblock \bibinfo{journal}{Nature} \bibinfo{volume}{471}
  (\bibinfo{year}{2011}) \bibinfo{pages}{83--86}.
\bibitem[{Koenig et~al.(2007)Koenig, Wiedmann, Bruene, Roth, Buhmann,
  Molenkamp, Qi, and Zhang}]{bib26}
\bibinfo{author}{M.~Koenig}, \bibinfo{author}{S.~Wiedmann},
  \bibinfo{author}{C.~Bruene}, \bibinfo{author}{A.~Roth},
  \bibinfo{author}{H.~Buhmann}, \bibinfo{author}{L.~W. Molenkamp},
  \bibinfo{author}{X.-L. Qi}, \bibinfo{author}{S.-C. Zhang},
\newblock \bibinfo{title}{Quantum spin {Hall} insulator state in hgte quantum
  wells},
\newblock \bibinfo{journal}{Science} \bibinfo{volume}{318}
  (\bibinfo{year}{2007}) \bibinfo{pages}{766--770}.
\bibitem[{Kato et~al.(2004)Kato, Myers, Gossard, and Awschalom}]{bib27}
\bibinfo{author}{Y.~Kato}, \bibinfo{author}{R.~Myers},
  \bibinfo{author}{A.~Gossard}, \bibinfo{author}{D.~Awschalom},
\newblock \bibinfo{title}{Observation of the spin {Hall} effect in
  semiconductors},
\newblock \bibinfo{journal}{Science} \bibinfo{volume}{306}
  (\bibinfo{year}{2004}) \bibinfo{pages}{1910--1913}.
\bibitem[{Kane and Mele(2005)}]{bib28}
\bibinfo{author}{C.~L. Kane}, \bibinfo{author}{E.~J. Mele},
\newblock \bibinfo{title}{${Z}_{2}$ topological order and the quantum spin
  {Hall} effect},
\newblock \bibinfo{journal}{Phys. Rev. Lett.} \bibinfo{volume}{95}
  (\bibinfo{year}{2005}) \bibinfo{pages}{146802}.
\bibitem[{Hsieh et~al.(2008)Hsieh, Qian, Wray, Xia, Hor, Cava, and
  Hasan}]{bib29}
\bibinfo{author}{D.~Hsieh}, \bibinfo{author}{D.~Qian},
  \bibinfo{author}{L.~Wray}, \bibinfo{author}{Y.~Xia}, \bibinfo{author}{Y.~S.
  Hor}, \bibinfo{author}{R.~J. Cava}, \bibinfo{author}{M.~Z. Hasan},
\newblock \bibinfo{title}{A topological {Dirac} insulator in a quantum spin
  {Hall} phase},
\newblock \bibinfo{journal}{Nature} \bibinfo{volume}{452}
  (\bibinfo{year}{2008}) \bibinfo{pages}{970--974}.
\bibitem[{Zhang et~al.(2020)Zhang, Han, Liao, Ye, and Liu}]{bib33}
\bibinfo{author}{T.~Zhang}, \bibinfo{author}{W.~Han}, \bibinfo{author}{R.-Y.
  Liao}, \bibinfo{author}{J.-W. Ye}, \bibinfo{author}{W.-M. Liu},
\newblock \bibinfo{title}{Supersolid phase of cold atoms},
\newblock \bibinfo{journal}{Eur. Phys. J. D} \bibinfo{volume}{74}
  (\bibinfo{year}{2020}) \bibinfo{pages}{138}.
\bibitem[{Martone and Stringari(2021)}]{bib34}
\bibinfo{author}{G.~I. Martone}, \bibinfo{author}{S.~Stringari},
\newblock \bibinfo{title}{{Supersolid phase of a spin-orbit-coupled
  {Bose}-{Einstein} condensate: A perturbation approach}},
\newblock \bibinfo{journal}{SciPost Phys.} \bibinfo{volume}{11}
  (\bibinfo{year}{2021}) \bibinfo{pages}{092}.
\bibitem[{Lyu et~al.(2024)Lyu, Chen, Zhu, and Zhang}]{bib35}
\bibinfo{author}{H.~Lyu}, \bibinfo{author}{Y.~Chen}, \bibinfo{author}{Q.~Zhu},
  \bibinfo{author}{Y.~Zhang},
\newblock \bibinfo{title}{Supercurrent-carrying supersolid in
  spin-orbit-coupled {Bose}-{Einstein} condensates},
\newblock \bibinfo{journal}{Phys. Rev. Res.} \bibinfo{volume}{6}
  (\bibinfo{year}{2024}) \bibinfo{pages}{023048}.
\bibitem[{Han et~al.(2018)Han, Zhang, Wang, Jiang, Zhang, and Zhang}]{bib30}
\bibinfo{author}{W.~Han}, \bibinfo{author}{X.-F. Zhang}, \bibinfo{author}{D.-S.
  Wang}, \bibinfo{author}{H.-F. Jiang}, \bibinfo{author}{W.~Zhang},
  \bibinfo{author}{S.-G. Zhang},
\newblock \bibinfo{title}{Chiral supersolid in spin-orbit-coupled {Bose} gases
  with soft-core long-range interactions},
\newblock \bibinfo{journal}{Phys. Rev. Lett.} \bibinfo{volume}{121}
  (\bibinfo{year}{2018}) \bibinfo{pages}{030404}.
\bibitem[{Liao(2018)}]{bib31}
\bibinfo{author}{R.~Liao},
\newblock \bibinfo{title}{Searching for supersolidity in ultracold atomic
  {Bose} condensates with rashba spin-orbit coupling},
\newblock \bibinfo{journal}{Phys. Rev. Lett.} \bibinfo{volume}{120}
  (\bibinfo{year}{2018}) \bibinfo{pages}{140403}.
\bibitem[{Han et~al.(2015)Han, Gediminas, Zhang, and Liu}]{bib32}
\bibinfo{author}{W.~Han}, \bibinfo{author}{J.~Gediminas},
  \bibinfo{author}{W.~Zhang}, \bibinfo{author}{W.-M. Liu},
\newblock \bibinfo{title}{Supersolid with nontrivial topological spin textures
  in spin-orbit-coupled {Bose} gases},
\newblock \bibinfo{journal}{Phys. Rev. A} \bibinfo{volume}{91}
  (\bibinfo{year}{2015}) \bibinfo{pages}{013607}.
\bibitem[{Liebster et~al.(2025)Liebster, Sparn, Kath, Duchene, Strobel, and
  Oberthaler}]{supersolid4}
\bibinfo{author}{N.~Liebster}, \bibinfo{author}{M.~Sparn},
  \bibinfo{author}{E.~Kath}, \bibinfo{author}{J.~Duchene},
  \bibinfo{author}{H.~Strobel}, \bibinfo{author}{M.~K. Oberthaler},
\newblock \bibinfo{title}{Supersolid-like sound modes in a driven quantum gas},
\newblock \bibinfo{journal}{Nat. Phys} \bibinfo{volume}{21}
  (\bibinfo{year}{2025}) \bibinfo{pages}{287–293}.
\bibitem[{Li et~al.(2017)Li, Lee, Huang, Burchesky, Shteynas, Top, Jamison, and
  Ketterle}]{bib36.5}
\bibinfo{author}{J.~Li}, \bibinfo{author}{J.~Lee}, \bibinfo{author}{W.~Huang},
  \bibinfo{author}{S.~Burchesky}, \bibinfo{author}{B.~Shteynas},
  \bibinfo{author}{F.~C. Top}, \bibinfo{author}{A.~O. Jamison},
  \bibinfo{author}{W.~Ketterle},
\newblock \bibinfo{title}{A stripe phase with supersolid properties in
  spin-orbit-coupled {Bose}-{Einstein} condensates},
\newblock \bibinfo{journal}{Nature} \bibinfo{volume}{543}
  (\bibinfo{year}{2017}) \bibinfo{pages}{91--93}.
\bibitem[{Zhang et~al.(2022)Zhang, Liu, and Zhang}]{bib36}
\bibinfo{author}{H.~Zhang}, \bibinfo{author}{S.~Liu}, \bibinfo{author}{Y.-S.
  Zhang},
\newblock \bibinfo{title}{Faraday patterns in spin-orbit-coupled
  {Bose}-{Einstein} condensates},
\newblock \bibinfo{journal}{Phys. Rev. A} \bibinfo{volume}{105}
  (\bibinfo{year}{2022}) \bibinfo{pages}{063319}.
\bibitem[{Wang et~al.(2025)Wang, Wang, Li, Dalfovo, and Qu}]{meiling2025}
\bibinfo{author}{M.~Wang}, \bibinfo{author}{J.~Wang}, \bibinfo{author}{Y.~Li},
  \bibinfo{author}{F.~Dalfovo}, \bibinfo{author}{C.~Qu},
\newblock \bibinfo{title}{Parametric excitations in a harmonically trapped
  binary {Bose}-{Einstein} condensate},
\newblock \bibinfo{journal}{Phys. Rev. A} \bibinfo{volume}{112}
  (\bibinfo{year}{2025}) \bibinfo{pages}{063303}.
\bibitem[{Liang et~al.(2024)Liang, Wang, Wang, and Li}]{bib52}
\bibinfo{author}{H.~Liang}, \bibinfo{author}{M.~Wang},
  \bibinfo{author}{J.~Wang}, \bibinfo{author}{Y.~Li},
\newblock \bibinfo{title}{Spin {Faraday} waves in periodically modulated
  spin-orbit-coupled {Bose} gases},
\newblock \bibinfo{journal}{Phys. Lett. A} \bibinfo{volume}{512}
  (\bibinfo{year}{2024}) \bibinfo{pages}{129592}.
\bibitem[{Liu and Wang(2024)}]{bib41.1}
\bibinfo{author}{X.~Liu}, \bibinfo{author}{X.~Wang},
\newblock \bibinfo{title}{Polygonal patterns of {Faraday} water waves analogous
  to collective excitations in {Bose}-{Einstein} condensates},
\newblock \bibinfo{journal}{Nat. Phys} \bibinfo{volume}{20}
  (\bibinfo{year}{2024}) \bibinfo{pages}{287–293}.
\bibitem[{Kwon et~al.(2021)Kwon, Mukherjee, Huh, Kim, Mistakidis, Maity,
  Kevrekidis, Majumder, Schmelcher, and Choi}]{bib41.2}
\bibinfo{author}{K.~Kwon}, \bibinfo{author}{K.~Mukherjee},
  \bibinfo{author}{S.~J. Huh}, \bibinfo{author}{K.~Kim}, \bibinfo{author}{S.~I.
  Mistakidis}, \bibinfo{author}{D.~K. Maity}, \bibinfo{author}{P.~G.
  Kevrekidis}, \bibinfo{author}{S.~Majumder}, \bibinfo{author}{P.~Schmelcher},
  \bibinfo{author}{J.-Y. Choi},
\newblock \bibinfo{title}{Spontaneous formation of star-shaped surface patterns
  in a driven {Bose}-{Einstein} condensate},
\newblock \bibinfo{journal}{Phys. Rev. Lett.} \bibinfo{volume}{127}
  (\bibinfo{year}{2021}) \bibinfo{pages}{113001}.
\bibitem[{Li et~al.(2012)Li, Pitaevskii, and Stringari}]{bib42}
\bibinfo{author}{Y.~Li}, \bibinfo{author}{L.~P. Pitaevskii},
  \bibinfo{author}{S.~Stringari},
\newblock \bibinfo{title}{Quantum tricriticality and phase transitions in
  spin-orbit coupled {Bose}-{Einstein} condensates},
\newblock \bibinfo{journal}{Phys. Rev. Lett.} \bibinfo{volume}{108}
  (\bibinfo{year}{2012}) \bibinfo{pages}{225301}.
\bibitem[{Inouye et~al.(1998)Inouye, Andrews, Stenger, Miesner, Stamper-Kurn,
  and Ketterle}]{bib38}
\bibinfo{author}{S.~Inouye}, \bibinfo{author}{M.~Andrews},
  \bibinfo{author}{J.~Stenger}, \bibinfo{author}{H.~Miesner},
  \bibinfo{author}{D.~Stamper-Kurn}, \bibinfo{author}{W.~Ketterle},
\newblock \bibinfo{title}{Observation of {Feshbach} resonances in a
  {Bose}-{Einstein} condensate},
\newblock \bibinfo{journal}{Nature} \bibinfo{volume}{392}
  (\bibinfo{year}{1998}) \bibinfo{pages}{151--154}.
\bibitem[{Pollack et~al.(2010)Pollack, Dries, Hulet, Magalh\~aes, Henn, Ramos,
  Caracanhas, and Bagnato}]{bib40}
\bibinfo{author}{S.~E. Pollack}, \bibinfo{author}{D.~Dries},
  \bibinfo{author}{R.~G. Hulet}, \bibinfo{author}{K.~M.~F. Magalh\~aes},
  \bibinfo{author}{E.~A.~L. Henn}, \bibinfo{author}{E.~R.~F. Ramos},
  \bibinfo{author}{M.~A. Caracanhas}, \bibinfo{author}{V.~S. Bagnato},
\newblock \bibinfo{title}{Collective excitation of a {Bose}-{Einstein}
  condensate by modulation of the atomic scattering length},
\newblock \bibinfo{journal}{Phys. Rev. A} \bibinfo{volume}{81}
  (\bibinfo{year}{2010}) \bibinfo{pages}{053627}.
\bibitem[{Moerdijk et~al.(1995)Moerdijk, Verhaar, and
  Axelsson}]{PhysRevA.51.4852}
\bibinfo{author}{A.~J. Moerdijk}, \bibinfo{author}{B.~J. Verhaar},
  \bibinfo{author}{A.~Axelsson},
\newblock \bibinfo{title}{Resonances in ultracold collisions of
  $^{6}\mathrm{Li}$, $^{7}\mathrm{Li}$, and $^{23}\mathrm{Na}$},
\newblock \bibinfo{journal}{Phys. Rev. A} \bibinfo{volume}{51}
  (\bibinfo{year}{1995}) \bibinfo{pages}{4852--4861}.
\bibitem[{Olshanii(1998)}]{PhysRevLett.81.938}
\bibinfo{author}{M.~Olshanii},
\newblock \bibinfo{title}{Atomic scattering in the presence of an external
  confinement and a gas of impenetrable bosons},
\newblock \bibinfo{journal}{Phys. Rev. Lett.} \bibinfo{volume}{81}
  (\bibinfo{year}{1998}) \bibinfo{pages}{938--941}.
\bibitem[{Cui et~al.(2010)Cui, Wang, and Zhou}]{PhysRevLett.104.153201}
\bibinfo{author}{X.~Cui}, \bibinfo{author}{Y.~Wang}, \bibinfo{author}{F.~Zhou},
\newblock \bibinfo{title}{Resonance scattering in optical lattices and
  molecules: Interband versus intraband effects},
\newblock \bibinfo{journal}{Phys. Rev. Lett.} \bibinfo{volume}{104}
  (\bibinfo{year}{2010}) \bibinfo{pages}{153201}.
\bibitem[{Eto et~al.(2016)Eto, Takahashi, Kunimi, Saito, and Hirano}]{bibfech2}
\bibinfo{author}{Y.~Eto}, \bibinfo{author}{M.~Takahashi},
  \bibinfo{author}{M.~Kunimi}, \bibinfo{author}{H.~Saito},
  \bibinfo{author}{T.~Hirano},
\newblock \bibinfo{title}{Nonequilibrium dynamics induced by
  miscible–immiscible transition in binary {Bose}–{Einstein} condensates},
\newblock \bibinfo{journal}{New Journal of Physics} \bibinfo{volume}{18}
  (\bibinfo{year}{2016}) \bibinfo{pages}{073029}.
\bibitem[{Chen et~al.(2019)Chen, Shibata, Eto, Hirano, and Saito}]{bib43}
\bibinfo{author}{T.~Chen}, \bibinfo{author}{K.~Shibata},
  \bibinfo{author}{Y.~Eto}, \bibinfo{author}{T.~Hirano},
  \bibinfo{author}{H.~Saito},
\newblock \bibinfo{title}{Faraday patterns generated by {Rabi} oscillation in a
  binary {Bose}-{Einstein} condensate},
\newblock \bibinfo{journal}{Phys. Rev. A} \bibinfo{volume}{100}
  (\bibinfo{year}{2019}) \bibinfo{pages}{063610}.
\bibitem[{Qu et~al.(2017)Qu, Pitaevskii, and Stringari}]{Qu_2017}
\bibinfo{author}{C.~Qu}, \bibinfo{author}{L.~P. Pitaevskii},
  \bibinfo{author}{S.~Stringari},
\newblock \bibinfo{title}{Spin–orbit-coupling induced localization in the
  expansion of an interacting {Bose}–{Einstein} condensate},
\newblock \bibinfo{journal}{New J. Phys.} \bibinfo{volume}{19}
  (\bibinfo{year}{2017}) \bibinfo{pages}{085006}.
\bibitem[{Nicolin(2011)}]{PhysRevE.84.056202}
\bibinfo{author}{A.~I. Nicolin},
\newblock \bibinfo{title}{Resonant wave formation in {Bose}-{Einstein}
  condensates},
\newblock \bibinfo{journal}{Phys. Rev. E} \bibinfo{volume}{84}
  (\bibinfo{year}{2011}) \bibinfo{pages}{056202}.
\bibitem[{Ticknor(2014)}]{node1}
\bibinfo{author}{C.~Ticknor},
\newblock \bibinfo{title}{Dispersion relation and excitation character of a
  two-component bose-einstein condensate},
\newblock \bibinfo{journal}{Phys. Rev. A} \bibinfo{volume}{89}
  (\bibinfo{year}{2014}) \bibinfo{pages}{053601}.
\bibitem[{Kagan et~al.(1996)Kagan, Surkov, and Shlyapnikov}]{bib48}
\bibinfo{author}{Y.~Kagan}, \bibinfo{author}{E.~L. Surkov},
  \bibinfo{author}{G.~V. Shlyapnikov},
\newblock \bibinfo{title}{Evolution of a {Bose}-condensed gas under variations
  of the confining potential},
\newblock \bibinfo{journal}{Phys. Rev. A} \bibinfo{volume}{54}
  (\bibinfo{year}{1996}) \bibinfo{pages}{R1753--R1756}.
\bibitem[{Wang et~al.(2024)Wang, Liang, Li, Li, and Qu}]{bib49}
\bibinfo{author}{J.~Wang}, \bibinfo{author}{H.~Liang}, \bibinfo{author}{Y.~Li},
  \bibinfo{author}{C.-H. Li}, \bibinfo{author}{C.~Qu},
\newblock \bibinfo{title}{Expansion dynamics of {Bose}-{Einstein} condensates
  in a synthetic magnetic field},
\newblock \bibinfo{journal}{Phys. Rev. A} \bibinfo{volume}{110}
  (\bibinfo{year}{2024}) \bibinfo{pages}{043307}.
\bibitem[{Lyvers and Mitchell(1988)}]{9114}
\bibinfo{author}{E.~Lyvers}, \bibinfo{author}{O.~Mitchell},
\newblock \bibinfo{title}{Precision edge contrast and orientation estimation},
\newblock \bibinfo{journal}{IEEE Trans. Pattern Anal. Mach. Intell.}
  \bibinfo{volume}{10} (\bibinfo{year}{1988}) \bibinfo{pages}{927--937}.

\end{thebibliography}

\end{document}